\documentclass[prl, longbibliography, twocolumn]{revtex4-1}
\usepackage{bm}
\usepackage{graphicx}
\usepackage{amsmath}
\usepackage{amssymb} 
\usepackage[utf8]{inputenc}
\usepackage[T1]{fontenc}
\usepackage{color}
\usepackage{xcolor}
\usepackage{upgreek}
\usepackage{subfigure}
\usepackage[unicode=true,colorlinks=true,citecolor=blue]{hyperref}
\usepackage{placeins}
\usepackage{lipsum}
\usepackage{epsfig}
\usepackage{wrapfig}
\usepackage[normalem]{ulem}
\usepackage{units}
\usepackage{cancel}
\usepackage{float}

\newcommand{\nix}[1]{}

\newcommand{\pderiv}[2]{\frac{\partial #1}{\partial #2}}
\renewcommand{\phi}{\varphi}

\renewcommand{\i}{\mathrm i}
\newcommand{\e}{\mathrm e}
\newcommand{\eps}{\varepsilon}

\newcommand{\beq}{\begin{equation}}
\newcommand{\eeq}{\end{equation}}
\newcommand\beqa{\begin{eqnarray}}
\newcommand\eeqa{\end{eqnarray}}
\newcommand\ba{\begin{array}}
\newcommand\ea{\end{array}}

\renewcommand\d{\partial}
 \newcommand{\aver}[1]{\left \langle #1 \right \rangle}

\begin{document}

\title{Second harmonic generation due to spatial structure of radiation beam}

\author{A. A. Gunyaga}
\affiliation{Ioffe Institute, 194021 St. Petersburg, Russia}

\author{M. V. Durnev} 
\email{durnev@mail.ioffe.ru}
\affiliation{Ioffe Institute, 194021 St. Petersburg, Russia}

\author{S. A. Tarasenko}
\affiliation{Ioffe Institute, 194021 St. Petersburg, Russia}

\begin{abstract}
We show that spatially structured radiation generates second harmonic even in homogeneous and isotropic media. The effect originates from non-locality of electric response to structured electromagnetic field.
We develop an analytical theory of such a second harmonic generation 
in conducting two-dimensional systems. For the general type of structured radiation, we
calculate the emerging electric currents at the double frequency and the emitted second harmonic radiation. 
The theory applied to twisted light reveals that the angular momentum of light doubles in the second harmonic emission.  Our results pave the way for light second harmonic generation and structuring beyond the constraints imposed by crystal symmetry.
\end{abstract}

%\pacs{
%}
\maketitle

%*************************************************************
%*************************************************************
\textit{Introduction.}---Second harmonic generation (SHG), a process in which two photons at the fundamental frequency 
$\omega$ merge into a photon with the double frequency $2\omega$, is the foundation stone of nonlinear physics~\cite{Bloembergen:1996,Hendry:2010,Butet:2015,Zhang:2020}.
It belongs to the class of second-order nonlinear phenomena and occurs in the leading (electric dipole) electron-photon interaction in noncentrosymmetric media.
Therefore, besides many applications in laser physics, SHG provides a powerful tool to study the symmetry properties of 
three-dimensional (3D) and two-dimensional (2D) crystalline materials~\cite{Dean:2009vu,Bykov:2013,Li:2013ui,Mennel:2018,Zhou:2020,Stepanov:2020,Wang:2009,Galanty:2018,Rubano:2019} as well as molecules and biological structures~\cite{Aghigh:2023,Asadipour:2024}.
Moreover, it has been recently realized that SHG and, in general, non-linear wave mixing allow one to create light beams with controllable structure of electromagnetic field~\cite{Buono:2022,Silva:2022,Ceglia:2024a,Meng:2020a, Zhang:2015,Dorrah:2022,Melik-Gaykazyan:2019}. The prominent examples of such a structured radiation are the vector beams composed of photons with different polarizations~\cite{Nesterov2000,Maurer2007} or the beams of twisted photons carrying orbital angular momenta~\cite{MolinaTerriza2007,Wei:2015,Choporova:2017,Knyazev2018,Forbes2021}. Recent experimental and theoretical research has revealed that structured 
electromagnetic field, also in infrared and terahertz ranges, gives rise to 
dc currents, which are sensitive to the local gradients of the Stokes polarization parameters and the orbital angular momentum of photons~\cite{Ji2020,Sederberg2020,Gunyaga:2023}. The beams of twisted photons are also promising to generate and probe skyrmions and other excitations in magnetic systems~\cite{Fujita:2017,Sirenko:2019,Sirenko:2021}, spiral waves in superconductors~\cite{Mizushima:2023}, and vortex states in atomic condensates~\cite{Andersen:2006}, to investigate excited states of atoms~\cite{Schmiegelow:2016}, and to manipulate micro- and nanoscale dielectric particles~\cite{Khonina:2004,Yang2021}.

The previous studies of SGH and the related non-linear light structuring were focused on systems with broken space inversion such as noncentrosymmetric crystals and metasurfaces. The studied second-order nonlinearity in 2D systems originates from
the inversion asymmetry of crystal lattice including 
the special stacks of 2D crystals and twisted van der Waals heterostructures~\cite{Hsu:2014,Yang:2020,Yao:2021,Zhang:2022,Paradisanos:2022},
the symmetry breaking by in-plane electric current~\cite{Khurgin:1995,Ruzicka:2012,Bykov:2012} or in-plane photon wavevector at oblique incidence of radiation~\cite{Glazov:2011,Mikhailov:2011},
the valley polarization~\cite{Golub:2014,Wehling:2015,Mouchliadis:2021}, 
and the edge effects in finite-size systems~\cite{Yin:2014,Mishina:2015,Durnev2022}. On the other hand, it is natural to expect that SHG emerges also in situation when inversion asymmetry is brought about by the radiation itself , as in the case of structured radiation with spatially inhomogeneous parameters.

\begin{figure}[ht]
    \centering
    \includegraphics[width=0.6\linewidth]{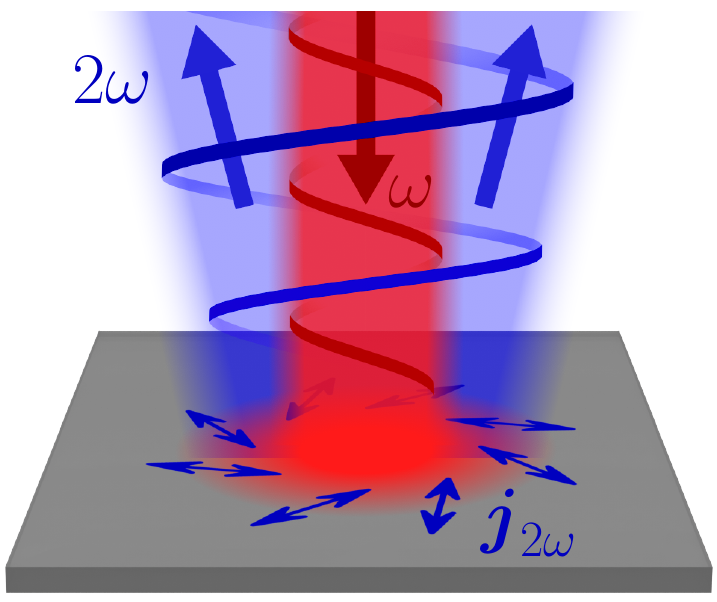}
    \caption{
   Second harmonic generation due to spatial structure of radiation. Inhomogeneous electric field of incident twisted radiation (incoming red beam) 
    oscillating at the fundamental frequency $\omega$ induces electric currents at the double frequency $\bm j_{2\omega}$ (blue arrows in the 
    2D plane). These currents, in turn, emit twisted electromagnetic field at the frequency $2\omega$ (outgoing blue beam).
    }
    \label{fig:scheme}
\end{figure}

Here, we demonstrate the SHG due to the radiation structure, Fig.~\ref{fig:scheme}. We consider the second-order response of an isotropic and homogeneous conducting 2D system to the structured radiation of a general type. As opposed to previously explored mechanisms, the second-order nonlinearity is driven here by non-local electronic response to the spatially varying high-frequency electromagnetic field. 
We develop an analytical theory of the SHG for intraband electron transport, which typically corresponds to microwave and terahertz  ranges for 2D semiconductors and up to optical and ultraviolet ranges for 2D metals. The calculations of non-linear currents are carried out in the framework of the quasi-classical Boltzmann approach similar to that used for the study of the edge SHG~\cite{Durnev2022} and edge photogalvanic effect~\cite{Karch2011,Candussio2020,Durnev2023} in conducting 2D systems and previously for the surface SHG in bulk metals~\cite{Jha:1965, 
Bloembergen:1968,Rudnick:1971,Sipe:1980}.
We apply the 
developed general theory to the vector beams and beams of twisted photons and show that the SHG can be used to double the orbital angular momentum.
The results indicate that the use of structured radiation expands the boundaries of non-linear optics.

%*************************************************************
%*************************************************************
\textit{Kinetic theory.}---Consider a structure with a two-dimensional electron gas (2DEG) in the $xy$ plane irradiated by a spatially inhomogeneous monochromatic electromagnetic wave, see Fig.~\ref{fig:scheme}. The electric field of the radiation acting upon the electrons has the form $\bm E(\bm r,t) = \bm E(r)\mathrm{e}^{-\i\omega t} + \mathrm{c.c.}$, where $\bm E(\bm r)$ is the spatially varying amplitude, and $\bm r = (x,y)$ is the in-plane coordinate. 
To the first order in the field amplitude, the electric field induces the currents at the same frequency $\omega$, which determine 
the linear response. The currents at the double frequency $\bm j_{2\omega}(\bm r, t) = \bm j_{2\omega}(\bm r)\mathrm{e}^{- 2\i \omega t} + \mathrm{c.c.}$, which emit the second harmonic radiation, appear in the second order response to the inhomogeneous electromagnetic field.

We calculate the currents $\bm j_{2\omega}(\bm r)$ by solving the quasi-classical Boltzmann equation 
\begin{equation}
\label{Boltzmann}
    \pderiv{f}{t} + \bm v\cdot\nabla f + e\left[\bm E_\parallel (\bm r,t) + \frac{\bm v}{c}\times\bm B (\bm r,t)\right] \cdot \pderiv{f}{\bm p} = I\{f\}
\end{equation}
for the distribution function $f(\bm p, \bm r ,t)$. Here, $\bm p = (p_x,p_y)$ and $\bm v = \bm p/m^*$ are the electron momentum and velocity, respectively, $e$ is the electron charge, $m^*$ is the effective mass, $\bm E_\parallel (\bm r,t)$ is the in-plane component of the actual electric field experienced by 2D electrons, $\bm B (\bm r,t)$ is the magnetic field, and $I\{ f\}$ is the collision integral. Since the electric field varies across the 2DEG plane, the electromagnetic wave necessarily contains the out-of-plane component of the magnetic field $B_z(\bm r,t) = B_z(\bm r) \mathrm{e}^{-\i\omega t} + \mathrm{c.c.}$ with $B_z = -\i (c/\omega) \left(\partial E_y/\partial x - \partial E_x/\partial y \right)$, which acts upon the moving electrons and, therefore, is included in Eq.~\eqref{Boltzmann}. The quasi-classical Eq.~\eqref{Boltzmann} is valid for the frequency range $\omega \ll E_F /\hbar$, where $E_F$ is the Fermi energy.

The distribution function is expanded in the series of powers of the electric field 
\begin{multline}
    f (\bm p, \bm r, t) = f_0 + \bigl[f_1 (\bm p, \bm r)\,\mathrm{e}^{-\i\omega t}+\mathrm{c.c.}\bigr]  \\
    + \bigl[f_2 (\bm p, \bm r) \,\mathrm{e}^{-2\i\omega t}+\mathrm{c.c.}\bigr],
\end{multline}
where $f_0(\eps)$ is the equilibrium distribution function with $\eps = p^2/2m^*$ being the electron energy, $f_1 \propto E$ is the first-order correction, and $f_2 \propto EE,\,EB$ is the second-order correction oscillating at $2\omega$. It is the correction $f_2$ that determines the local density of the electric current at $2\omega$ 
\begin{equation}
\label{j2w_0}
    \bm j_{2\omega} (\bm r) = e\nu\sum\limits_{\bm p}\bm vf_2 (\bm p, \bm r),
\end{equation}
where $\nu$ is the factor of spin/valley degeneracy.

We solve the kinetic Eq.~\eqref{Boltzmann} and calculate the current~\eqref{j2w_0} 
for the collision integral in the form $I\{f\} = -(f - \aver{f})/\tau$, where $\tau$ is the relaxation time and $\aver{f}$ is the distribution function averaged over the directions of $\bm p$. It is assumed that the field $\bm E_\parallel$ varies smoothly in the 2DEG plane, so that the scale of the field inhomogeneity $L$ is much larger than the mean free path of electrons $l = v_F \tau$ and the length $v_F/\omega$, where $v_F$ is the Fermi velocity. In this case, the gradients of the field components and the distribution function can be treated perturbatively as small corrections. It is also assumed that $L \gg (2\pi \sigma/c) \lambda$, where $\sigma$ is the 2D conductivity, and $\lambda = 2\pi c/\omega$ is the wavelength of the incident field, which allows us to neglect the screening of inhomogeneous field by 2DEG.

Calculation of the current at the double frequency Eq.~\eqref{j2w_0} yields~\footnote{See Supplemental Material, which includes Ref.~\cite{Bateman:1954}, for details on derivation of Eqs.~\eqref{general_current} and~\eqref{j2w_q}.} \nocite{Bateman:1954}
\begin{multline}\label{general_current}
    \bm j_{2\omega} = \frac{-\i e\sigma_0\tau}{m^*\omega (1 - \i\omega \tau)(1-2\i\omega\tau)^2}\biggl[(1-\i\omega\tau)\nabla\bigl(\bm E_\parallel\cdot\bm E_\parallel \bigr) \\- \bigl(\bm E_\parallel\cdot\nabla\bigr)\bm E_\parallel + (1-4\i\omega\tau)\bm E_\parallel\bigl(\nabla\cdot\bm E_\parallel\bigr)\biggr]\:,
\end{multline}
where $\sigma_0 = n e^2 \tau/m^*$ is the static 2DEG conductivity, and $n = \nu \sum_{\bm p} f_0$ is the 2D electron density.
Equation~\eqref{general_current} is the main result of our paper. It describes the currents at the double frequency emerging 
in 2D conducting systems in response to inhomogeneous electromagnetic field with a general spatial structure.

\begin{figure}
    \centering    \includegraphics[width=\linewidth]{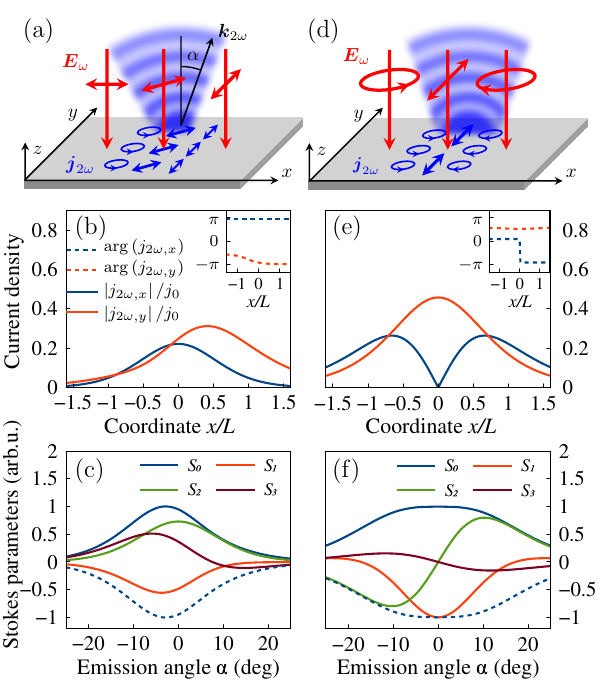}
    \caption{Electric currents at double frequency induced by incident radiation
    with spatially inhomogeneous polarization and the corresponding second harmonic emission. 
    (a) Sketch of the current distribution $\bm j_{2\omega}$ and the emitted second harmonic radiation (outgoing blue waves) for linearly polarized incident field $\bm E_{\omega}$ whose polarization vector varies in the 2D plane. (b) Spatial distribution of the $x$ and $y$ components of the current density $\bm j_{2\omega}$; the main graph and the inset show the absolute values and the arguments of the complex-value currents, respectively.
    (c) Stokes parameters of the emitted radiation at the frequency $2\omega$ as functions of the emission angle $\alpha$. The dashed curve shows the mirrored $S_0(\alpha)$ dependence. (d)-(f) The same for the incident radiation with the circular polarization varying from left-handed to right-handed. The results are presented for $\omega \tau = 1$ and $k_{2\omega}L=4\pi$.}
    \label{fig:1D_gradients}
\end{figure}

%*************************************************************
%*************************************************************
\textit{Polarization driven SHG.}---Now we apply the general Eq.~\eqref{general_current} to study the SHG induced by electromagnetic field
with the constant in-plane amplitude $|\bm E_\parallel|$ but non-uniform polarization varying along some axis (here, the $x$ axis).  
We consider two examples of a such field which are sketched in Fig~\ref{fig:1D_gradients}. 

In the first example, Fig.~\ref{fig:1D_gradients}(a), the field is linearly polarized and its polarization rotates 
from $\bm E \parallel x$ at large negative $x$ to $\bm E \parallel y$ at large positive $x$. 
We take the spatial profile of the field amplitude in the form $\bm E_\parallel (x) = E_0 [\cos\Phi_1(x/L), \sin\Phi_1(x/L)]$,
where $\Phi_1$ is a smooth function varying from $\Phi_1(-\infty) = 0$ to $\Phi_1 (+\infty)= \pi/2$ and $L$ is the scale where 
the polarization varies. Since the product $\bm E_\parallel \cdot \bm E_\parallel = E_0^2$ is constant, the first term in Eq.~\eqref{general_current} 
vanishes. The other two terms give the local density of electric current at the double frequency
\begin{eqnarray}
\label{j2w_lin}
    j_{2\omega,x} (x)  &=& \frac{2\omega\tau j_0 \sin 2 \Phi_1 (x/L)}{(1-\i\omega\tau)(1-2\i\omega\tau)^2}\, \Phi_1'(x/L) \:, \\
    j_{2\omega,y} (x) &=& \frac{\i j_0 [1 - 4\i\omega \tau \sin^2 \Phi_1(x/L)]}{(1-\i\omega\tau)(1-2\i\omega\tau)^2}\, \Phi_1'(x/L) \:,\nonumber
\end{eqnarray}
where
\begin{equation}
\label{j0}
    j_0 = \frac{ne^3\tau^2E_0^2}{{m^*}^2\omega L}\:.
\end{equation}
At low frequencies, $\omega \tau \ll 1$, the current $\bm j_{2\omega}$ flows along $y$. 
In the high-frequency limit, $\omega \tau \gg 1$, the current $\bm j_{2\omega}$ is also linearly polarized but its 
polarization varies from $\bm j_{2\omega} \parallel x$ to $\bm j_{2\omega} \parallel y$. 
The spatial distribution of the current density $\bm j_{2\omega} (x)$ for the intermediate case $\omega \tau = 1$ and 
$\Phi_1(\xi) = (\pi/4)[\tanh(\xi) + 1]$ is shown in Fig.~\ref{fig:1D_gradients}(b). The current $\bm j_{2\omega}$  has both 
 $x$ and $y$ components and is generally elliptically polarized. The phase shift between the components depends on $x$
 and so does the degree of ellipticity. 
 
In the second example, Fig.~\ref{fig:1D_gradients}(d), the polarization of incident field varies from the left-handed to the right-handed circular polarization through the linear polarization. The field profile is taken in the form $\bm E_\parallel (x) = E_0 [\i \sin \Phi_2(x/L), \cos \Phi_2(x/L)]$ with the function 
$\Phi_2$ varying from  $\Phi_2 (- \infty)= -\pi/4$ to $\Phi_2 (+\infty) = \pi/4$. The corresponding current density obtained from Eq.~\eqref{general_current} 
is
\begin{eqnarray}
\label{j2w_circ}
  j_{2\omega,x} (x)  &=& \frac{2\i j_0 \sin 2 \Phi_2 (x/L)}{(1-\i\omega\tau)(1-2\i\omega\tau)}\,\Phi_2'(x/L)\:, \\
    j_{2\omega,y} (x) 
    &=&  \frac{j_0 [1 - 4\i\omega \tau \cos^2 \Phi_2(x/L)]}{(1-\i\omega\tau)(1-2\i\omega\tau)^2}\,\Phi_2'(x/L) \:. \nonumber
\end{eqnarray} 
The current $\bm j_{2\omega}$ is linearly polarized along $y$ at $x=0$ and elliptically polarized everywhere else, see Fig.~\ref{fig:1D_gradients}(d).
The spatial distributions of the $x$ and $y$ components of $\bm j_{2\omega}$ and the phase shift between them are shown 
in Fig.~\ref{fig:1D_gradients}(e) for $\omega \tau =1$ and $\Phi_2(\xi) = (\pi/4)\tanh(\xi)$.

The ac current $\bm j_{2\omega}(x, t) = \bm j_{2\omega}(x)\mathrm{e}^{- 2\i \omega t} + \mathrm{c.c.}$ in the 2DEG plane emits electromagnetic waves with the vector potential $\bm A_{2\omega} (x, z, t) = \bm A_{2\omega} (x, z) \e^{-2\i\omega t} + \mathrm{c.c}$. The field $\bm A_{2\omega} (x, z)$ can be found from the wave equation~\cite{Landau_vol2}
\begin{equation}\label{Helmholtz} 
    \Delta\bm A_{2\omega} + k_{2\omega}^2\bm A_{2\omega} = -\frac{4\pi}{c}\bm j_{2\omega}(x)\delta(z)\:,
\end{equation}
where $k_{2\omega} = 2\omega/c$ and $\delta(z)$ is the Dirac delta function. 

Far from the region where the current is generated, $\bm A_{2\omega} (x, z)$ represents the outgoing cylindrical wave.
The  amplitude and polarization of the wave depend on the emission angle $\alpha$ counted from the 2D normal, $\tan \alpha = x/z$, see Fig.~\ref{fig:1D_gradients}(a). By using the Green function of the 2D Helmholtz equation (the Hankel function of the first kind), we solve Eq.~\eqref{Helmholtz} and find the asymptotic of $\bm A_{2\omega} (x, z)$ at large distances $R = \sqrt{x^2 + z^2} \gg L , \lambda$
\begin{multline}
\label{A1}
    \bm A_{2\omega} (x, z) = \frac{\i \sqrt{2\pi}}{c \sqrt{k_{2\omega}R}} \exp \left[\i\left(k_{2\omega} R - \frac{\pi}{4} \right) \right] \\
    \times \int \bm j_{2\omega}(x') \exp \left( -\i k_{2\omega, x} x' \right) dx'\:,
\end{multline}
where $k_{2\omega,x} = k_{2\omega} \sin \alpha$ is the $x$ component of the wave vector $\bm k_{2\omega} = (x/R, 0 , z/R) k_{2\omega}$, Fig.~\ref{fig:1D_gradients}(a). The angular distribution of the field $\bm A_{2\omega} (x, z)$ is determined by the Fourier image of the current
$ \bm j_{2\omega}(x)$.  The amplitudes of the emitted magnetic and electric fields at the double frequency are given by 
$\bm B_{2\omega} = \i \bm k_{2\omega} \times \bm A_{2\omega}$ and 
$\bm E_{2\omega} = \bm B_{2\omega} \times \bm k_{2\omega} / k_{2\omega}$, respectively.  

To describe the angular distribution of the intensity and polarization of the second harmonic emission, we introduce the Stokes parameters   
\begin{eqnarray}
\label{Stokes}
    S_0 &=& |E_{2\omega,x'}|^2 + |E_{2\omega,y}|^2\:,~~S_1 = |E_{2\omega,x'}|^2 - |E_{2\omega,y}|^2\:, \nonumber \\
    S_2 &=& E_{2\omega,x'} E_{2\omega,y}^* + E_{2\omega,x'}^* E_{2\omega,y}\:, \nonumber \\
    S_3 &=& \i (E_{2\omega,x'} E_{2\omega,y}^* - E_{2\omega,x'}^* E_{2\omega,y} ) 
\end{eqnarray}
in the coordinate frame $(x', y, z')$, where $z'$ is parallel to the wave vector of the emitted wave $\bm k_{2\omega}$. 
The parameter $S_0$ determines the local intensity of the emitted radiation, $S_1$ and $S_2$ describe the linear polarization 
in the $(x', y)$ plane, and $S_3$ describes the circular polarization.

Figures~\ref{fig:1D_gradients}(c) and \ref{fig:1D_gradients}(f) present the Stokes parameters of the second harmonic field
as functions of the emission angle $\alpha$. The curves are calculated for the current distributions, which are given by Eqs.~\eqref{j2w_lin} and \eqref{j2w_circ} and shown in Fig.~\ref{fig:1D_gradients}(b) and \ref{fig:1D_gradients}(e), respectively, for $k_{2\omega} L = 4 \pi$.
The emitted signal is mainly focused within a sector defined by the angle $\Delta \alpha \sim 1/(k_{2\omega} L)$. 
As shown in Figs.~\ref{fig:1D_gradients}(c) and \ref{fig:1D_gradients}(f), the polarization of the second harmonic wave depends 
significantly on the emission direction. In particular, in Fig.~\ref{fig:1D_gradients}(f), the wave emitted upwards is polarized along the $y$ axis ($S_1/S_0 = -1$) whereas with increasing $|\alpha|$ the linear polarization rotates in the $(x',y)$ plane approaching $\pm \pi/4$ orientation ($S_2/S_0 = \pm 1$) at large $|\alpha|$. 

Interestingly, in Fig.~\ref{fig:1D_gradients}(c), the peak intensity is emitted at a nonzero angle. This is due to the phase variation of $j_{2\omega,y}$, see the inset of Fig.~\ref{fig:1D_gradients}(b), which results in the shift of the Fourier image center from $k_{2\omega, x} = 0$. Note also a large degree of circular polarization ($S_3/S_0$ ratio) of the emission at $\alpha \leq 0$, which results from the phase shift between the $x$ and $y$ components of the current.

To conclude, we estimate the power of the second harmonic emission. At $k_{2\omega} L \gg 1$,
the radiation is emitted within a narrow sector $\Delta \alpha \sim 1/(k_{2\omega} L)$, and the power emitted from the length $L_y$ of the sample 
can be estimated as $P_{2\omega} / L_y \propto c R S_0 (0) \Delta \alpha$, where $S_0(0)$ is the Stokes parameter for the upward emission.
Using the estimation $ \int \bm j_{2\omega}(x)  dx  \sim j_0 L$, we obtain $P_{2\omega} / L_y \propto j_0^2 L /c$ with 
$j_0$ given by Eq.~\eqref{j0}. 
For 2D structure with the carrier density $n = 10^{12}$~cm$^{-2}$, 
the momentum relaxation time $\tau = 1$~ps, the effective mass $m^* = 0.03m_0$, which corresponds to bilayer graphene~\cite{Candussio2020}, and the incident radiation with $\omega \tau =1$
and $E_0 = 0.25$~kV/cm, corresponding to terahertz radiation with the intensity $I = 1$~kW/cm$^2$, and $L = 2\pi c/\omega$, 
we obtain $j_0 \approx 0.2$~mA/cm and the emitted power $P_{2\omega} /L_y$ of the order of 0.2~$\mu$W/cm. In a single-layer graphene, one may expect even higher values due to the smaller effective mass.

The above estimation corresponds to the second-order nonlinear susceptibility $\chi^{(2)} \sim 10^5$~nm$^2$/V in the terahertz range. Projection of this estimation to the infrared range with $\hbar \omega = 0.1$~eV gives $\chi^{(2)} \sim 0.1$~nm$^2$/V. The nonlinear susceptibility of the same order of magnitude is measured in twisted bilayer graphene~\cite{Yang:2020} and MoS$_2$ and WS$_2$ monolayers~\cite{Kumar:2013,Malard:2013} due to the lack of crystal lattice inversion symmetry.

%*************************************************************
%*************************************************************
\textit{SHG induced by twisted light.}---Now we apply the developed theory to twisted radiation, as shown in Fig.~\ref{fig:scheme}. As an example, we consider the class of non-diffracting Bessel beams characterized by the total angular momentum projection $m$
~\cite{Knyazev2018, Maurer2007,Sederberg2020,Nesterov2000}. 
Electric field of the Bessel beam is given by~\cite{Knyazev2018,Gunyaga:2023}
\begin{equation}\label{Bessel_electric}
\boldsymbol{E}(\bm r,z) = E_0 \mathrm{e}^{\i k_z z} 
\sum\limits_{\bm k_{\parallel}} a( k_{\parallel}) \exp \left(\mathrm{i}\bm k_{\parallel}\cdot\bm r + \i m \phi_k \right) \bm e_{\bm k}\:,
\end{equation}
where $a(k_{\parallel}) = \i^{-m-1}  (2\pi/ k_{\parallel})  \delta(k_{\parallel} - k_0 )$.
It is a superposition of the plane waves with the wave vectors $\bm k = (\bm k_{\parallel},k_z)$ where  
$|\bm k_\parallel | = k_0$ and $|\bm k| = \omega/c$ are fixed, i.e., $\bm k$ is lying on the surface of the cone with the angle $\theta_k = \arcsin k_0 /k $. 
The plane waves are shifted in phase as determined by the exponent $\e^{\i m \phi_k}$, where 
$\phi_k$ in the polar angle of the wave vector $\bm k_\parallel$.
Polarization of each wave is determined by the unit vector $\bm e_{\bm k} \perp \bm k$ decomposed as $\bm e_{\bm k} = \alpha \, \bm e_{\theta k} + \beta \, \bm e_{\varphi k}$, where $\bm e_{\theta k}$ and $\bm e_{\varphi k}$ are the local orthogonal unit vectors in the directions of increasing $\theta_k$ and $\phi_k$, respectively, and $|\alpha|^2 + |\beta|^2 = 1$.

The Fourier components 
of the emerging in-plane current at the double frequency $\bm j_{2\omega} (\bm r) = \sum_{\bm q}\bm j_{2\omega, \bm q} \exp(\i \bm q \cdot \bm r)$ 
in the paraxial approximation $\theta_k \ll 1$ read~\cite{Note1}
\begin{multline}
\label{j2w_q}
    \bm j_{2\omega, \bm q} = -\frac{ n e^3 \tau^2 E_0^2 \exp(2\i m \phi_q)}{{m^*}^2 q\omega (1 - \i \omega \tau)(1-2\i\omega\tau)^2}  b(q) \times \\
    \left( a_{2\omega} \, \bm q + b_{2\omega} \bm e_z \times \bm q \right)   \,.
\end{multline}
Here, 
%$\bm n_{q}=\bm k_{\parallel}/k_{\parallel}$, 
$\bm e_z$ is the unit vector parallel to $z$, $\varphi_q$ is the polar angle of $\bm q$,
\begin{equation}
\label{b_k}
    b(q) = \frac{(-1)^m\Theta(2k_0 - q)}{\sqrt{4k_0^2 - q^2}} \:,
\end{equation}
$\Theta$ is the Heaviside function, and
\begin{eqnarray}
\label{alpha_beta}
a_{2\omega} &=&(\alpha^2 + \beta^2) (1 - 2\i\omega \tau) (q/k_0)^2 - 4(\alpha^2 - \beta^2) \i \omega \tau\:, \nonumber \\
b_{2\omega} &=& -8 \i \omega \tau \alpha \beta.    
\end{eqnarray}
The current density in the real space $\bm j_{2\omega} (\bm r)$, given by the Fourier transform of Eq.~\eqref{j2w_q}, is proportional to $k_0$, which determines the scale of the rings in the Bessel beam cross section $L \propto 1/k_0$. Thus, $\bm j_{2\omega} (\bm r) \propto 1/L$, as expected.

The electric field of the second harmonic radiation emitted by the current~\eqref{j2w_q} in the half-space $z>0$ is given by
\begin{multline}
\label{E2w}
\bm E_{2\omega}(\bm r, z) =  \frac{2\pi n e^3\tau^2 E_0^2}{{m^*}^2 c \omega  (1 - \i \omega \tau)(1-2\i\omega\tau)^2}  \times\\  \sum\limits_{\bm k_{\parallel}} b(k_{\parallel})  \exp\left(\i\bm k_{\parallel} \cdot \bm r + 2\i m \phi_k + \i k_{z} z \right) \zeta \bm e_{\bm k}\:, 
\end{multline}
where $k_{z} = \sqrt{k_{2\omega}^2 - k_\parallel^2}$, $\bm e_{\bm k} = \alpha_{2\omega} \bm e_{\theta k} + \beta_{2\omega} \bm e_{\varphi k}$,
$\alpha_{2\omega} = a_{2\omega}/ \zeta$, $\beta_{2\omega} = b_{2\omega}/\zeta$, and $\zeta = \sqrt{|a_{2\omega}^2|+|b_{2\omega}^2|}$.

Equation~\eqref{E2w} describes the emitted second harmonic field $\bm E_{2\omega}$ induced by the incident Bessel beam with the angular momentum $m$. 
The emitted beam is also twisted with the doubled angular momentum $2 m$.

\begin{figure}
    \centering
    \includegraphics[width=1\linewidth]{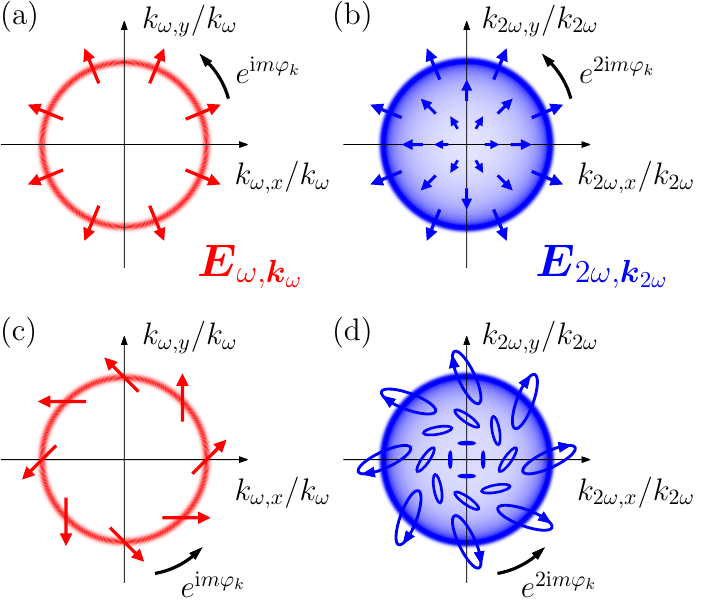}
    \caption{
    Fourier images of the incident Bessel beams at the frequency $\omega$ (a, c) and the corresponding emitted beams at the frequency $2\omega$ (b, d). Red and blue gradients encode the amplitude  of the incident and emitted waves, respectively. Arrows show the polarization. The distributions are plotted for radially polarized incident beam (a, b) and diagonally polarized incident beam (c, d).
    }
    \label{fig:Bessel-beam}
\end{figure}

Figure~\ref{fig:Bessel-beam} shows the examples of amplitude and polarization patterns of the incident Bessel beams at the fundamental frequency and the corresponding emitted beams at the double frequency. The Bessel beams are constructed from the plane waves with the fixed in-plane wave vectors $|\bm k_\parallel| = k_0$. Therefore, their images in the Fourier space  are rings, Figs~\ref{fig:Bessel-beam}(a) and ~\ref{fig:Bessel-beam}(c). 
In contrast to the incident waves, the emitted waves are superpositions of the Bessel beams with the frequency $2\omega$ and the in-plane wave vectors $|\bm k_{\parallel}|$ lying in the range from 0 to $2k_0$, see Eq.~\eqref{E2w}.
Their images in the Fourier space are shown in Figs.~\ref{fig:Bessel-beam}(b) and~\ref{fig:Bessel-beam}(d) by the blue gradients. The images are determined by the dependence $\zeta b(k_\parallel)$, which gives the weights of individual Bessel beams in the emitted wave.
The fact that the in-plane wave vectors of the second harmonic waves lie within the circle $|\bm k_\parallel| = 2 k_0$ stems from the conservation of the in-plane momentum in the SHG process when 
two photons with $\bm k_\parallel^{(1)}$ and $\bm k_\parallel^{(2)}$ merge into the photon with 
$\bm k_\parallel^{(1)} + \bm k_\parallel^{(2)}$.
The diverging amplitude of the emitted waves at $|\bm k_\parallel| = 2 k_0$ originates from the diverging 
density of 
states which can be composed from $\bm k_\parallel^{(1)}$ and $\bm k_\parallel^{(2)}$ so that $|\bm k_\parallel^{(1)} + \bm k_\parallel^{(2)}| \approx 2 k_0$.

Polarization of the waves constituting the incident and emitted beams is sketched in Fig.~\ref{fig:Bessel-beam} by red and blue arrows. It is determined by the polarization vector $\bm e_{\bm k}$, i.e., by the parameters $\alpha$ and $\beta$ for the incident beams and $\alpha_{2\omega}$ and $\beta_{2\omega}$ for the emitted beams, respectively. 
In particular, radially polarized ($\alpha = 1$, $\beta = 0$), Fig.~\ref{fig:Bessel-beam}(a), and azimuthally polarized ($\alpha = 0$, $\beta = 1$) Bessel beams generate second harmonic beams that are both radially polarized, $|\alpha_{2\omega}| = 1$, $\beta_{2\omega} = 0$, Fig.~\ref{fig:Bessel-beam}(b). The incident beam composed of circularly polarized plane waves ($\alpha = 1/\sqrt{2}$, $\beta = \pm \i/\sqrt{2}$) generates the beam of circularly polarized waves with the same helicity, $\beta_{2\omega} = \pm \i \alpha_{2\omega}$. The most peculiar polarization pattern of the second harmonic emission is induced by the incident field that is diagonally linearly polarized, e.g., $\alpha = \beta = 1/\sqrt{2}$, Fig.~\ref{fig:Bessel-beam}(c). In that case, both $\alpha_{2\omega}$ and $\beta_{2\omega}$ are nonzero and their relative phase depends on the emission wave vector $\bm k_\parallel$, Fig.~\ref{fig:Bessel-beam}(d).

%*************************************************************
%*************************************************************
\textit{Conclusions.}---To summarize, we have studied the second harmonic generation (SHG) of spatially structured radiation in centrosymmetric two-dimensional systems.
The inversion asymmetry required for the SHG is brought about by the structure of electric and magnetic fields in the incident radiation beam.
 We have obtained analytical expressions for the currents and emission at the double frequency and applied the results to examine the polarization driven SHG and the SHG induced by twisted radiation. In the latter case, the second harmonic emission is also twisted with the doubled angular momentum and, in general case, has a complex polarization pattern. Our results pave the way for second
harmonic generation and structuring in two-dimensional materials beyond the constraints imposed by crystal
lattice.

\acknowledgments 
This work was supported by the Russian Science Foundation (Project No. 22-12-00211).

\bibliographystyle{apsrev4-1-customized}
\bibliography{bibliography}

%*****************************************************************
%*****************************************************************
%*****************************************************************
%Supplementary material
\onecolumngrid
\appendix
\newpage

\widetext
\begin{center}
\textbf{\large Second harmonic generation due to spatial structure of radiation beam: Supplementary material}
\end{center}

\section{Derivation of the general expression for the current at the double frequency}

In this section, we derive Eq.~(4) of the main text, which gives the general expression for the current density $\bm j_{2\omega}$.
The functions $f_1$ and $f_2$ given by Eq.~(2) of the main text satisfy the following differential equations, which follow from Eq. (1),
\begin{align}
	-\i\omega f_1 + \bm v\cdot\nabla f_1 + e\bm E_{\parallel} (\bm r) \cdot\pderiv{f_0}{\bm p} = I\{f_1\}\:,\label{pert1}
 \\[1ex]
	-2\i\omega f_2 + \bm v\cdot\nabla f_2 + e\left[\bm E_{\parallel}(\bm r)+\frac{\bm v}{c}\times\bm B (\bm r)\right]\cdot\pderiv{f_1}{\bm p} = I\{f_2\}\:.\label{pert2}
\end{align}
The term with the Lorentz force in Eq.~\eqref{pert1} vanishes because $\d f_0/\d \bm p \propto \bm p$ and $\bm v \times \bm B$ is orthogonal to $\bm p$.
Multiplying Eq.~\eqref{pert2} by $\bm v$, summing up over $\bm p$ and integrating by parts, one obtains
\begin{equation}\label{current_non_final}
\bm j_{2\omega} = -\frac{e \nu \tau}{1-2\i\omega\tau}\left[\sum\limits_{\bm p}\bm v(\bm v\cdot\!\nabla f_2)-\frac{e \bm E_{\parallel}(\bm r)}{m^*} \sum\limits_{\bm p}f_1 + \frac{e \bm B(\bm r)}{m^* c}\times\sum\limits_{\bm p}\bm v f_1\right].
\end{equation}

For smoothly varying field, the gradients of the field components and the distribution function are small and can be treated perturbatively. The third term in Eq.~\eqref{current_non_final} contains $\bm B(\bm r)$, which is already proportional to the gradients of electric field; therefore, the sum $\sum_{\bm p} \bm v f_1$ can be calculated in the local response approximation. Multiplying Eq.~\eqref{pert1} by $\bm v$, summing up the resulting equation over $\bm p$, and neglecting the gradient term $\propto \nabla f_1$, one obtains
\begin{equation}
\label{jw}
   e\nu \sum\limits_{\bm p}\bm v f_1=\sigma \bm E_{\parallel}(\bm r)\:,
\end{equation}
where $\sigma = \sigma_0/(1-\i\omega\tau)$ is the high-frequency conductivity.
The sum $\sum_{\bm p} f_1$ in the second term in Eq.~\eqref{current_non_final} can be calculated by summing up Eq.~\eqref{pert1} over $\bm p$ and using Eq.~\eqref{jw}, which yields
\begin{equation}
\label{rhow}
   e\nu \sum\limits_{\bm p}f_1 = -\frac{\i\sigma}{\omega}\,\nabla \cdot \bm E_{\parallel}(\bm r)\:,
\end{equation}
where $\nabla \cdot \bm E_{\parallel} = \partial E_x/\partial x + \partial E_y/\partial y$.

The first term in Eq.~\eqref{current_non_final} is proportional to the gradient of $f_2$; therefore, the sums $\sum_{\bm p} v_\alpha v_\beta f_2$ can also be calculated in the local response approximation. These sums are determined by the zero and second angular harmonics of $f_2$. Multiplying Eq.~\eqref{pert2} by $v_{x}^2 + v_{y}^2$, $v_x v_y$ or $v_x^2 - v_y^2$ and summing up the result over $\bm p$, one obtains
\begin{equation}
\label{v2f2}
    \nu\sum\limits_{\bm p}\frac{v_x^2+v_y^2}{2}f_2 = \frac{\i\sigma}{2m^*\omega} \bm E_\parallel\cdot\bm E_\parallel \:,
\end{equation}
and
\begin{eqnarray}
\label{vxvyf2}
\nu\sum\limits_{\bm p}v_xv_yf_2 &=& \frac{2\tau\sigma}{m^*(1-2\i\omega\tau)} E_{x}E_{y}\:, \nonumber \\
    \nu\sum\limits_{\bm p}\frac{v_x^2 - v_y^2}{2}f_2 &=& \frac{\tau\sigma}{m^*(1-2\i\omega\tau)}\left(E_{x}^2 - E_{y}^2\right)\:.
\end{eqnarray}
In the derivation of Eq.~\eqref{v2f2}, we assumed that $\omega \tau_\eps \gg 1$, where $\tau_\eps$ is the energy relaxation time.

Finally, substituting Eqs.~\eqref{jw}-\eqref{vxvyf2} into Eq.~\eqref{current_non_final} for the current, using $B_z = -\i(c/\omega)(\partial E_y/\partial x - \partial E_x/\partial y)$ and regrouping the terms, we obtain Eq.~(4) of the main text.

\section{Fourier component of the current density induced by the Bessel beam}

Here, we derive Eq.~(12) of the main text for the Fourier component
\begin{equation}
\bm j_{2\omega,\bm q} = \int\bm j_{2\omega}(\bm r)\exp(-\i\bm q \cdot \bm r)\,\mathrm{d}\bm r
\end{equation}
of the current at the double frequency induced by the Bessel beam. The current density $\bm j_{2\omega}(\bm r)$ is found from the general Eq.~(4) with account for the electric field of the Bessel beam Eq.~(11).

We start by calculating the Fourier transform of the first term in Eq.~(4):
\begin{equation}\label{first}
\int\nabla\bigl(\bm E_\parallel\cdot\bm E_\parallel \bigr)\e^{-\i\bm q \cdot \bm r}\,\mathrm{d}\bm r = \i E_0^2(2\pi)^2\sum\limits_{\bm q_1}\sum\limits_{\bm q_2}(\bm q_1+\bm q_2)a(q_1)a(q_2)\e^{\i m(\varphi_{q1}+\varphi_{q2})}\delta(\bm q_1+\bm q_2-\bm q)(\bm e_{q1}\cdot\bm e_{q2})\:,
\end{equation}
where $\bm q_{1}$ and $\bm q_2$ are the in-plane wave vectors with the polar angles $\phi_{q1}$ and $\phi_{q2}$, $\bm e_{q1} \perp \bm q_1$ and $\bm e_{q2} \perp \bm q_2$ are the polarization vectors, and $\delta$ is the Dirac delta function. Delta function in Eq.~\eqref{first} poses a restriction $\bm q_1 + \bm q_2 = \bm q$. Taking also into account that $|\bm q_1|=|\bm q_2| = k_0$ [which follows from $a(q_j) \propto \delta(q_j-k_0)$], one obtains $\e^{\i m(\varphi_{q1}+\varphi_{q2})} = \e^{2\i m \varphi_q}$, where $\phi_q$ is the polar angle of $\bm q$, and hence
\begin{equation}
   \int\nabla\bigl(\bm E_\parallel\cdot\bm E_\parallel \bigr)\e^{-\i\bm q \cdot \bm r}\,\mathrm{d}\bm r = \i (-1)^{m+1} \frac{E_0^2 \bm q \e^{2\i m\varphi_{q}}}{k_0^2}
 \iint\mathrm{d}\bm q_1\mathrm{d}\bm q_2\delta(q_1 - k_0)\delta(q_2-k_0)\delta(\bm q_1+\bm q_2 - \bm q)\,(\bm e_{q1}\cdot\bm e_{q2})\:.
\end{equation}

In the paraxial approximation ($\theta_k \ll 1$) the polarization vector $\bm e_{\bm k} \approx \alpha \bm n_k + \beta (\bm e_z \times \bm n_k)$, where $\bm n_k = \bm k_{\parallel}/k_0$ and $\bm e_z$ is the unit vector parallel to $z$.
Therefore, the scalar product of polarization vectors is
\begin{equation}
    \bm e_{q1}\cdot\bm e_{q2} \approx (\alpha^2+\beta^2) \frac{\bm q_1\cdot\bm q_2}{k_0^2}\:.
\end{equation}
Using this equation and taking into account that $\bm q_1\cdot\bm q_2 = (q^2 - 2k_0^2)/2$, one finally obtains
\begin{equation}
\label{first_final}
   \int\nabla\bigl(\bm E_\parallel\cdot\bm E_\parallel \bigr)\e^{-\i\bm q \cdot \bm r}\,\mathrm{d}\bm r = \i (-1)^{m+1} \frac{E_0^2 \e^{2\i m\varphi_{q}}}{2 k_0^4}(\alpha^2+\beta^2) (q^2 - 2k_0^2) \bm q\Delta(q)\:.
\end{equation}
Here, we introduced the following integral
\begin{equation}
\label{delta}
    \Delta(q) = \iint\mathrm{d}\bm q_1\mathrm{d}\bm q_2\delta(q_1 - k_0)\delta(q_2-k_0)\delta(\bm q_1+\bm q_2 - \bm q)\:.
\end{equation}
This integral can be written in the form $\Delta(q) = 2 \pi k_0^2 \int  r J_0(q r) J_0^2(k_0 r) \mathrm{d} r$, which is essentially the Hankel transform of the function $J_0^2(k_0 r)$, resulting in~\cite{Bateman:1954}
\begin{equation}
\label{delta2}
    \Delta(q) = \frac{4k_0^2}{q \sqrt{4k_0^2 - q^2}} \Theta(2k_0 -q)\:.
\end{equation}
The Heaviside function in Eq.~\eqref{delta2} reflects the restriction $q \leq 2k_0$, which follows directly from $\bm q = \bm q_1 + \bm q_2$ and $|\bm q_1|=|\bm q_2| = k_0$.

Let us now compute  the Fourier transform of the second term in Eq.~(4):
\begin{equation}\label{second}
 \int \bigl(\bm E_\parallel\cdot \nabla \bigr) \bm E_\parallel \e^{-\i\bm q \cdot \bm r}\,\mathrm{d}\bm r = 
     \i (-1)^{m+1} \frac{E_0^2 \e^{2\i m\varphi_{q}}}{k_0^2} 
    \iint\mathrm{d}\bm q_1\mathrm{d}\bm q_2\delta(q_1 - k_0)\delta(q_2-k_0)\delta(\bm q_1+\bm q_2 - \bm q)\,(\bm e_{q1}\cdot \bm q_2) \bm e_{q2}\:.
\end{equation}
In the paraxial approximation
\begin{equation}
    \bigl(\bm e_{q1}\cdot\bm q_2\bigr)\bm e_{q2} \approx
\frac{\bm q_1\cdot \bm q_2}{k_0^2}\left[ \alpha^2 \bm q_2 +   \alpha\beta (\bm e_z\times\bm q_2)\right] + 
\frac{(\bm q_1\times\bm q_2)\cdot \bm e_{z}}{k_0^2} \left[ \alpha \beta \bm q_2 + \beta^2 (\bm e_z\times\bm q_2)  \right]\:.
\end{equation}
After integration with delta functions in Eq.~\eqref{second} the terms $\bigl(\bm q_1\cdot \bm q_2\bigr) \bm q_2$ and $\bigl[(\bm q_1\times\bm q_2)\cdot \bm e_{z}\bigr] \bm q_2$ yield
$\bm q (q^2 - 2k_0^2)/4$ and $(\bm e_z \times \bm q) (4k_0^2 - q^2)/4$, respectively, resulting in
\begin{equation}
\label{second_final}
 \int \bigl(\bm E_\parallel\cdot \nabla \bigr) \bm E_\parallel \e^{-\i\bm q \cdot \bm r}\,\mathrm{d}\bm r = 
     \i (-1)^{m+1} \frac{E_0^2 \e^{2\i m\varphi_{q}}}{4k_0^4} 
    \left[\alpha^2  (q^2 - 2k_0^2) \bm q + \beta^2 (q^2 - 4k_0^2) \bm q +2 \alpha \beta k_0^2 (\bm e_z \times \bm q)  \right] \Delta(q)\:.
\end{equation}

Finally, similar calculations for the third term in Eq.~(4) yield
\begin{equation}
\label{third_final}
 \int \bm E_\parallel \bigl(\bm E_\parallel\cdot \nabla \bigr) \e^{-\i\bm q \cdot \bm r}\,\mathrm{d}\bm r = 
     \i (-1)^{m+1} \frac{E_0^2 \e^{2\i m\varphi_{q}}}{2k_0^2}
   \left[\alpha^2 \bm q + \alpha \beta (\bm e_z \times \bm q)  \right] \Delta(q)\:.
\end{equation}

Substituting Fourier transforms of the contributions to the current Eqs.~\eqref{first_final}, \eqref{second_final} and \eqref{third_final} into Eq.~(4), we finally obtain Eq.~(12) of the main text.

\end{document}